\documentclass[12pt]{article}
\usepackage{latexsym}
\usepackage{amssymb}
\usepackage[dvips]{graphicx}
\usepackage{epsfig}
\usepackage{rotating}

\usepackage{color}
\usepackage{cancel}

\usepackage{amsfonts}
\usepackage{amsmath}

\newcommand{\bea}{\begin{eqnarray}}
\newcommand{\eea}{\end{eqnarray}}
\newcommand{\beq}{\begin{equation}}
\newcommand{\eeq}{\end{equation}}

\def\v{\vskip.2cm}
\makeatletter

\begin{document}

\begin{titlepage}

\begin{flushright}
pi-math-ph-280\\
ICMPA-MPA/2012/010\\
\end{flushright}

\begin{center}

{\Large\bf Enhanced quantization on the circle}

\bigskip

Joseph Ben Geloun$^{a,c,*}$
 and John R. Klauder$^{b,\dag}$

\bigskip

$^a${\em Perimeter Institute for Theoretical Physics, 31 Caroline
St N} \\
{\em ON N2L 2Y5, Waterloo, ON, Canada} \\
\medskip
$^{b}${\em Department of Physics and
Department of Mathematics}\\
{\em University of Florida, Gainesville, FL 32611-8440}\\
\medskip
$^{c}${\em International Chair in Mathematical Physics
and Applications}\\
{\em ICMPA--UNESCO Chair, 072 B.P. 50  Cotonou, Republic of Benin} \\
\medskip
E-mails:  $^{*}$jbengeloun@perimeterinstitute.ca,
\quad $^{\dag}$klauder@phys.ufl.edu

\begin{abstract}
We apply the quantization scheme introduced in
[arXiv:1204.2870] to a particle on a circle. We find that the quantum action functional restricted to appropriate coherent states can be expressed as the classical action plus  $\hbar$-corrections. This result extends the examples presented in the cited paper.
\end{abstract}

\end{center}

\noindent Pacs numbers: 03.65.-w

\noindent  Key words: Quantization, coherent states, periodic coordinates.

\vspace{0.5cm}
\begin{center}
\today
\end{center}
\end{titlepage}

\setcounter{footnote}{0}

\section{Introduction}

Conventional canonical quantization promotes a classical momentum
$p$ into a Hermitian operator $P$ and a classical position $q$
into a Hermitian operator $Q$ which obey $[Q,P]=i\hbar$.
Other classical quantities such as the Hamiltonian find
a quantum counterpart as
$H(p,q) \longrightarrow \mathcal{H} = H(P,Q)$  plus
possible corrections of $O(\hbar)$. This prescription
works very well for many systems but it also has its limitations.

Enhanced quantization \cite{Klauder:2012vh}\cite{Klauderbook} offers a new interpretation
of the very process of quantization that encompasses the usual
canonical story and offers additional features as well.
Provided both $P$ and $Q$ are self-adjoint operators
(a stronger condition than Hermiticity) we can generate
unitary operators  acting on a fiducial state $|\eta\rangle$
to yield
\beq
|p,q\rangle  =  e^{-\frac{i}{\hbar}qP} e^{\frac{i}{\hbar}pQ}|\eta\rangle
\eeq
as a set of coherent states which spans the Hilbert space
$\mathfrak{H}$. If the quantum action functional
\beq
A_Q = \int_0^T \, \langle \psi(t)|\, \big[ i\hbar\partial_t - \mathcal{H}\big]
\,|  \psi(t)\rangle\,  dt
\eeq
is used in a variational principle, it leads to the Schr\"odinger equation
$\big[ i\hbar\partial_t - \mathcal{H}\big]|  \psi(t)\rangle =0$ as
an equation of motion. However, arbitrary variations of  $|  \psi(t)\rangle$
for a microscopic system are not accessible to a macroscopic observer
who can only change the velocity or position of the microscopic
system leading to the fact that the only state she/he could make
are those represented by $|p(t),q(t)\rangle$.
The restricted quantum action functional becomes
\bea
A_{Q(R)}\hskip-1.4em &&= \int_0^T \, \langle p(t),q(t)|\, \big[ i\hbar\partial_t - \mathcal{H}\big]
\,|  p(t),q(t)\rangle \, dt\crcr
&&= \int_0^T \, \Big[ p(t)\dot{q}(t)-H(p(t),q(t))\Big] dt
\eea
which, because $\hbar>0$ still, may be called an enhanced
classical action functional whereas the usual classical action is given by
\beq
A_C = \int_0^T \, \Big[ p(t)\dot{q}(t)-H_c(p(t),q(t))\Big] dt\,,
\eeq
 where $H_c(p(t),q(t)) = \lim_{\hbar\to 0} H(p(t),q(t))$.
In this view, then, classical theory is a subset of quantum
theory, and they both co-exist just like they do in the real world where
$\hbar >0$ \cite{Klauder:2012vh}.

Other two-dimensional  continuous sheets of unit
vectors may also correspond to a classical  canonical
system, and in \cite{Klauder:2012vh}, two other sets
of coherent states were shown to have this property.
In this paper, we consider the enhanced quantization
of a classical particle moving on a circle of finite radius.
This system leads to yet another set of coherent states which
serves to unite the classical and quantum theories for such a system.
This is our main result.

Quantum mechanics on a nontrivial topological space has been
discussed in many contexts and still attracts attention (see \cite{Govaerts:1999ep,Govaerts:2006uu,ChagasFilho:2008ud}
and more references therein) and
coherent states on the circle have been also discussed in
\cite{al,Gazeau:2009zz,Kowalski:1998hx,Gonzalez:1998kj,chad}. Nevertheless, none of these contributions
addresses the issue of the  relationship between classical and quantum
actions as presented here. This paper
provides another instance for which the rationale introduced in \cite{Klauder:2012vh} yields an interesting quantum/classical connection.

\section{Enhanced quantization  on the circle}

Consider a particle on the circle $S^1$. The position space
can be parametrized by $\theta \in [-\pi,\pi)$.
We consider, at the quantum level, a set of quantum operators
$Q$ and $P$, associated with position and momentum,
 satisfying the commutation relation:
\beq
[Q,P] = i \hbar\,.
\label{heis}
\eeq
The spectrum of the operator $Q$ is bounded in $[-\pi,\pi)$ and, like
$\theta$, is periodic.
To start the program of \cite{Klauder:2012vh},
the operators $P$ and $Q$ have to be self-adjoint such that the operators  $e^{-\frac{i}{\hbar} q P}$ and $e^{-\frac{i}{\hbar} p Q}$ keep
their ordinary and useful unitary feature. \v

\noindent{\bf Self-adjoint extension of $P$ on the circle.}
Self-adjoint extensions are well known on the space of square integrable functions on any finite interval $[A,B)$,
 with vanishing boundary conditions
(see for instance \cite{Reed:1975uy}). In a streamlined analysis, we review the properties of
$P=-i\hbar \partial_\theta$ as an operator acting on $L^2([-\pi,\pi), d\theta)$.

Let us investigate the domain $D(P)$ of $P$ for
$P$ being symmetric, i.e., $P^\dag = P$ on $D(P)$.
Consider the inner product for any two functions $\psi,\varphi,\psi',\varphi' \in  L^2([-\pi,\pi), d\theta)$ (with as yet unspecified boundary values) given by
\beq
(\psi, P \varphi)=(-i\hbar) \int_{-\pi}^{\pi} \; \psi^*(\theta)\; \varphi'(\theta)\;\; d\theta
= (-i\hbar) \psi^*  \varphi \Big|^{\pi}_{-\pi}
+ (P\psi, \varphi)\,,
\eeq
so that for $(\psi, P \varphi) = (P\psi, \varphi)$ to hold on $D(P)$, one requires
\beq
\psi^*(\pi)  \varphi(\pi) - \psi^*(-\pi)  \varphi(-\pi) =0\;.
\label{psiphi}
\eeq
This condition is satisfied if we adopt $\varphi(\pm \pi)= 0$ and make no restriction on $\psi$. In this case
$D(P) = \big\{\varphi;\; \varphi, \varphi' \in L^2([-\pi,\pi), d\theta); \;\;
\varphi(\pi) = \varphi(-\pi) =0 \big\}$.
However, the domain of $P^\dag$,
$D(P^\dag) = \big\{\varphi;\; \varphi, \varphi' \in L^2([-\pi,\pi), d\theta) \big\} \supset D(P)$.

Defining the self-adjoint extension of $P$ is a procedure aimed
at rendering $D(P^\dag) = D(P)$ by enlarging $D(P)$
and restricting $D(P^\dag)$ so they coincide \cite{Reed:1975uy}.
In particular, the condition \eqref{psiphi} can also be fulfilled  by
imposing the boundary condition, $\varphi(\pi) = e^{2\pi i\alpha} \varphi(-\pi)$,
for a given $\alpha\in [0,1)$, and thus we can enlarge the domain of $P$
and reduce the domain of $P^\dag$ so that
\beq
\widetilde{D}(P_\alpha) = \Big\{\varphi;\; \varphi, \varphi' \in L^2([-\pi,\pi), d\theta); \;\;
\varphi(\pi) = e^{2\pi i\alpha}\varphi(-\pi)\Big\} = \widetilde{D}(P_\alpha^\dag)
\,.\label{domp}
\eeq
As noticed in \cite{Klauder:2012vh}, having defined self-adjoint
extensions for $P_\alpha$ and $Q$ (which is trivial here), we can
define unitary operators by exponentiating these generators.\v

\noindent{\bf Coherent states on the circle.}
Let us now pursue the quantum program
associated with \eqref{heis} henceforth denoting $P$
by $P_\alpha$, where $\alpha\in[0,1)$ labels the
different inequivalent representations of the momentum
operator. We will use units such that
$Q$ is dimensionless
and so the dimension of $P$ is that of $\hbar$.
We define eigenvectors $ |\theta\rangle $ for
the operator $Q$, satisfying
$\langle \theta|\theta'\rangle=\delta_{S^1}(\theta-\theta')$,
where $\delta_{S^1}$ should be understood as periodic on $S^1$,
as well as eigenvectors $|n,\alpha \rangle$ of $P_\alpha$, obeying $\langle n,\alpha| m,\alpha\rangle=\delta_{n,m}$,
 such that
\beq
Q |\theta\rangle  = \theta|\theta\rangle  \,, \qquad
P_\alpha |n,\alpha \rangle = p_{n,\alpha}  |n,\alpha \rangle \,, \qquad
 \langle   \theta | P_\alpha |n,\alpha \rangle = (-i\hbar) \partial_{\theta}
\langle   \theta |n,\alpha \rangle  = p_{n,\alpha} \langle   \theta |n,\alpha \rangle\,.
\eeq
It is well known that the spectrum of $P_\alpha$ on the circle is such that
$p_{n,\alpha} =\hbar(n +\alpha)$, $(n,\alpha) \in \mathbb{Z}\times [0,1)$ (more features
introduced by the topology of the manifold and
 self-adjoint properties of $P$  can be found in \cite{Govaerts:1999ep}\cite{Govaerts:2006uu}\cite{ChagasFilho:2008ud}). As a realization
of the functions $ \langle   \theta |n,\alpha \rangle$,
we find  that the normalized wave functions  are given by
\beq
\langle   \theta |n,\alpha \rangle  = \frac{1}{\sqrt{2\pi}} e^{i (n + \alpha )\theta} \;.
\eeq
The  self-adjoint operators $P_\alpha$ and $Q$ yield
the unitary operators $e^{-\frac{i}{\hbar} q P_\alpha}$ and $e^{-\frac{i}{\hbar} p Q}$, where $(q,p)\in S^1 \times \mathbb{R}$. From these operators, a set of states is defined  by
\bea
|p,q\rangle  = e^{-\frac{i}{\hbar} q P_\alpha}e^{\frac{i}{\hbar} p Q}\;
 |\eta_\alpha\rangle\,,
\eea
where $|\eta_\alpha\rangle$ is called the fiducial state.
We verify that the set $\{|p,q\rangle\}$ satisfies

\begin{enumerate}
\item[(i)] A normalization condition:
$\langle  p,q|p,q\rangle  =\langle  \eta_\alpha |\eta_\alpha\rangle=1$,
since $\langle  \eta_\alpha |\eta_\alpha\rangle $ is normalized.
\item[(ii)] A resolution of unity:
\beq
\int_{\mathbb{R}\times S^1}  |p,q\rangle \langle  p,q|   \,\frac{dp dq}{2\pi\hbar}
 = I_{\mathfrak{H}}\,.
\label{resolu}
\eeq
Indeed, it can be  shown that, for all $\theta,\theta' \in [\pi,\pi)$,
\beq
\int_{\mathbb{R} \times S^1} \langle \theta |p,q\rangle \langle  p,q| \theta'\rangle  \frac{dp dq}{2\pi\hbar}
=
\delta_{S^1}(\theta-\theta')
\int_{S^1} \langle \theta - q | \eta_\alpha\rangle \langle \eta_\alpha| \theta-q \rangle   dq =
\langle \theta|\theta'\rangle \langle \eta_\alpha|\eta_\alpha\rangle\,,
\eeq
where we used the fact that $\int_{\theta-\pi}^{\theta+\pi}dq |q\rangle \langle q| = \int_{-\pi}^{\pi}dq |q\rangle \langle q|$ because of  periodicity.
 Thus \eqref{resolu} is recovered for a normalized $|\eta_\alpha\rangle$.
\end{enumerate}
The set of states $\{|p,q\rangle\}$ forms an overcomplete
family of normalized states.  Henceforth, these states will be called coherent states.

We can now discuss the dynamics associated with such states
by introducing a general quantum Hamiltonian of the form
${\mathcal H}(P, e^{i Q}, e^{-i Q})$.
It has been argued that quantum  and
classical mechanics should co-exist (see for instance
\cite{kbs}--\cite{howfar}). In the present situation
and using the coherent  states, we establish
a link between the quantum and the classical actions.

Consider the restricted quantum action associated with
$|\psi(t)\rangle\rightarrow|p(t),q(t)\rangle$ defined above,
which leads to
\beq
A_{Q(R)} = \int_0^T \;
\langle p(t),q(t)| \Big[i\hbar \partial_t\; - \;{\mathcal H} \Big]
| p(t),q(t) \rangle\; dt\,.
\eeq
As explained earlier, within this action functional
a macroscopic observer can vary, not
the entire Hilbert space of states, but only the coherent-state
subspace,  when studying a microscopic system.
We choose a class of fiducial vectors satisfying
\beq
\langle \eta_\alpha |Q |\eta_\alpha\rangle =0
\qquad \text{and} \qquad
\langle\eta_\alpha |P_\alpha |\eta_\alpha\rangle =\hbar \alpha
\label{qpmean}
\eeq
so that $|\eta_\alpha\rangle$ is in the domain of the self-adjoint operators $Q$ and $P_\alpha$.

In the ordinary canonical  situation, the choice of the fiducial vector $|\eta\rangle$ as the ground state of an harmonic oscillator
makes it an extremal weight vector: $(Q+iP)|\eta\rangle=0$, and the latter relation yields
  $\langle \eta|Q|\eta\rangle=0$ and
$\langle \eta|P|\eta\rangle=0$. Hence,  \eqref{qpmean}
can be considered  an analog of this condition
modified slightly due to the topology of the configuration space.

A straightforward calculation  using \eqref{qpmean} yields
\beq
\langle p(t),q(t)| \Big[i\hbar \partial_t \Big] | p(t),q(t) \rangle
= (\hbar\alpha + p)  \dot{q} \,.
\label{palph}
\eeq
Furthermore, we have
\beq
H_\alpha(p(t),q(t)) :=
\langle p(t),q(t)| \;{\mathcal H}(P_\alpha, e^{i Q}, e^{-i Q}) \;| p(t),q(t) \rangle
=  \langle\eta_\alpha |   {\mathcal H} (P_\alpha + p, e^{i (Q+q)}, e^{-i (Q+q)})
| \eta_\alpha \rangle \,.
\eeq
Hence, the restricted quantum action reads
\beq
A_{Q(R)} = \int_0^T\;
\Big\{
  \Big[  p  \dot{q} -
 H_\alpha(p(t),q(t))\Big]+\hbar\,\alpha \dot{q}\Big\}dt\,,
\eeq
where the term
$\widetilde{A}_{C} = \int_0^T[ p\dot{q}    - H_\alpha(p(t),q(t))  ]dt$,
 as in the ordinary situation \cite{Klauder:2012vh}, can be related to
a classical action $A_C$ up to $\hbar$ corrections since
\beq
 H_\alpha(p(t),q(t))  = H_{c,\alpha}(p,q) +O(\hbar;p,q)\,.
\eeq
In the last equation,  $H_{c,\alpha}(p,q)$ is viewed as the usual classical Hamiltonian.
Interestingly, we notice that the quantum parameter $\alpha$
induces a surface term $\hbar \alpha \dot{q} $ in $A_{Q(R)}$ which  makes no influence on the enhanced, or the classical, equations of motion whatsoever.  At the end, we can write
\bea
A_{Q(R)} = A_C + O(\hbar)\;.
\eea

Let us discuss, in more detail, a general case which can be evaluated completely. We choose the quantum Hamiltonian as
\beq
\mathcal{H} (P_\alpha,e^{iQ}, e^{-iQ}) = P_\alpha^2 + V(e^{iQ}, e^{-iQ})\,,
\qquad
 V(e^{iQ}, e^{-iQ})
= a_0+\sum_{n=1}^{m} \big[ a_n \cos n Q + b_n \sin n Q\big]\,,
\eeq
with mass units chosen so that $1/2\mu=1$ and
where $m\in \mathbb{N}$. Next, we choose a particular fiducial vector such that, for $0<b<1$,
\beq
\eta_\alpha(\theta)  := \langle \theta | \eta_\alpha\rangle = N \,e^{i\alpha \theta}\Big[1+b\cos\theta\Big]^{r/\hbar}\;,
\;
N = \Big[ 2\pi  \left[1-b\right]^{2r/\hbar}\,
   _2F_1\left(\frac{1}{2},-2
   \frac{r}{\hbar};1;-\frac{ 2b}{1-b}\right) \Big]^{-1/2},
\label{fidu}
\eeq
with $r/\hbar>m$ an integer, $_2F_1$ denotes the ordinary hypergeometric function \cite{abram} and $N$ is a normalization factor fixed such that $\langle \eta_\alpha | \eta_\alpha\rangle=1$. Note that $|\eta_\alpha(\theta)|$ is even and
periodic and that \eqref{qpmean} is satisfied.
For large $r/\hbar\gg1$,
the following approximation is valid:
\beq
|\eta_\alpha(\theta)|^2 =
|\langle \theta | \eta_\alpha\rangle|^2 = N^2
e^{\frac{2r}{\hbar}  [\ln\big(1+b\cos\theta\big) ]}
\leq  N^2
e^{\frac{2r}{\hbar}  [- \frac{b\theta^2}{2(1+b)} - \ln (1+b)]}
= \tilde N^2 e^{-\frac{r}{\hbar} \frac{b\theta^2}{(1+b)}  }\;;
\eeq
hence $|\theta|\lesssim \sqrt{\hbar/r}$ which is small.
One can therefore consider $\eta_\alpha(\theta)$ as cutting-off large
$\theta$-values.
Evaluating the diagonal coherent state matrix elements of $\mathcal{H}$, and setting $\alpha'=\hbar\alpha$, leads to
\bea
&&\hskip-2em\langle p(t),q(t)| \;{\mathcal H}(P_\alpha, e^{i Q}, e^{-i Q}) \;| p(t),q(t) \rangle
=  \langle\eta_\alpha | \;  (P_\alpha + p)^2
+ V(e^{i (Q+q)}, e^{-i (Q+q)}) \;
| \eta_\alpha \rangle\crcr
&&=
 (\alpha' + p)^2 - \alpha'^2 +
\langle\eta_\alpha | P^2 | \eta_\alpha \rangle
+
a_0+\sum_{n=1}^{m} \Big[ a_n \cos n q   + b_n \sin n q \Big]  + O(\hbar)\crcr
&&=   (p+\alpha' )^2  + V(e^{iq}, e^{-iq}) + O(\hbar)\;,
\eea
where $\langle\eta_\alpha |  P_\alpha^2 | \eta_\alpha \rangle $
and $\alpha'^2$ are constants included in $O(\hbar)$.
Thus,  we can infer that
\bea
A_{Q(R)} = \int_0^T \; \Bigg\{ (p + \alpha')\dot{q}
-\Big[  (p+\alpha')^2
+  V(e^{iq}, e^{-iq})\Big]  +O(\hbar)
\Bigg\}dt\,.
\eea
Therefore, up to constants and  a canonical shift in momentum ($p\rightarrow p  + \alpha'$),
\bea
A_{Q(R)} = \int_0^T \; \Bigg\{ p\dot{q}
-\Big[ p^2  +  V(e^{iq}, e^{-iq}) \Big]  +O(\hbar) \Bigg\}dt = A_{C}
+ O(\hbar)\,.
\eea

Canonical transformations are well defined in the present setting.
These transformations involve 
a change of variables
$(p,q) \to (\tilde p, \tilde q)$ such that the symplectic structure
at the classical level is preserved leading to $\{\tilde q, \tilde p\} = 1 = \{q,p\}$
as well as $pdq= \tilde p d\tilde q+d\tilde G(\tilde p,\tilde q)$,
where $\tilde G$ is the generator of such a transformation.
Coherent states in the transformed coordinates, namely $|\tilde p,\tilde q\rangle$, are chosen to be the same
as before the  change of variables.
Thus, it is clear that the restricted quantum action computed
with respect to $|\tilde p,\tilde q\rangle \equiv |p(\tilde p,\tilde q),q(\tilde p,\tilde  q)\rangle  = |p,q\rangle$ yields a
classical action $\tilde A_C$ corresponding to $A_C$
up to a surface term given by $\dot{\tilde G}(\tilde p, \tilde q)$. No change of the operators is involved. \v

\noindent{\bf Coherent state induced geometry on phase space.}
The geometry of the coherent states \cite{Klauder:2001rb} can be investigated  as well
using the Fubini-Study metric element,
$\widetilde c_\alpha\, ds_\alpha^2 = ||\, d|p,q\rangle ||^2  - |\langle p,q|d|p,q\rangle |^2$, given $\widetilde c_\alpha= c_\alpha /\hbar^2$.
We have $d|p,q\rangle  = -(i/\hbar) e^{-\frac{i}{\hbar} q P_\alpha} e^{\frac{i}{\hbar} p Q} \Big[  dq(P_\alpha + p )  -   dp Q\Big]
 |\eta_\alpha\rangle$, and therefore
\bea
||\, d|p,q\rangle ||^2
&=&
\frac{1}{\hbar^2}
    \Big[  dq^2 ( D'_{\alpha}(r)+2p\alpha' + p^2)  +  dp^2  D_{\alpha}(r)
\Big] \;,\crcr
|\langle p,q|d|p,q\rangle |^2  &=& \frac{1}{\hbar^2}
|\langle\eta_\alpha|\,\Big[  dq(P_\alpha + p )  -   dp Q\Big]
 |\eta_\alpha\rangle |^2 =
 \frac{1}{\hbar^2}  (p+ \alpha')^2 dq^2 \;,
\eea
where $D_\alpha(r):= \langle \eta_\alpha|Q^2 |\eta_\alpha \rangle$ and $D'_{\alpha}(r):= \langle \eta_\alpha|P^2 |\eta_\alpha \rangle$
are both  constants.
Finally, the metric on the subspace of coherent states can be written as
\beq
 c_\alpha ds_\alpha^2 = ( D'_{\alpha}(r)- \alpha'^2)  dq^2 +  D_{\alpha}(r)  dp^2\,, \qquad
ds_\alpha^2 =  A_\alpha dq^2 +  dp^2 = dq_\alpha^2 + dp^2\,,
\eeq
where we have set $q_\alpha = \sqrt{|A_\alpha|}q$, $c_\alpha =D_{\alpha}(r)$ and
$A_\alpha =  ( D'_{\alpha}(r) - \alpha'^2)/ D_{\alpha}(r) $. The metric $ds_\alpha^2 $ describes a flat geometry 
with a cylindrical topology. If desired, this metric can be imposed on the classical phase space as well.

\section*{Acknowledgements}
JRK thanks the Perimeter Institute, Waterloo, Canada, for its hospitality.
Research at Perimeter Institute is supported by the Government of Canada through
Industry Canada and by the Province of Ontario through the Ministry of Research and Innovation.


\begin{thebibliography}{99}

\bibitem{Klauder:2012vh}
  J.~R.~Klauder,
  ``Enhanced Quantization: A Primer,''
  arXiv:1204.2870 [quant-ph].

\bibitem{Klauderbook}
  J.~R.~Klauder,
{\it A Modern Approach to Functional Integration},
   (Birkh\"auser, Boston, MA, 2010).

\bibitem{Govaerts:1999ep}
  J.~Govaerts and V.~M.~Villanueva,
  ``Topology classes of flat U(1) bundles and diffeomorphic covariant representations of the Heisenberg algebra,''
  Int.\ J.\ Mod.\ Phys.\ A {\bf 15}, 4903 (2000)
  [quant-ph/9908014].

\bibitem{Govaerts:2006uu}
  J.~Govaerts and F.~Payen,
  ``Topological Background Fields as Quantum Degrees of Freedom of Compactified Strings,''
  Mod.\ Phys.\ Lett.\ A {\bf 22}, 119 (2007)
  [hep-th/0608023].

\bibitem{ChagasFilho:2008ud}
  W.~Chagas-Filho,
  ``2T Physics and Quantum Mechanics,''
  arXiv:0802.2840 [hep-th].

\bibitem{al}
S.~T.~Ali, J.-P.~Antoine and J.-P.~Gazeau,
{\it Coherent States, Wavelets, and their Generalizations},
(Springer-Verlag, Berlin,  2000).

\bibitem{Gazeau:2009zz}
  J.-P.~Gazeau,
  {\it Coherent states in quantum physics},
  (Wiley-VCH, Weinheim, 2009).


\bibitem{Kowalski:1998hx}
  K.~Kowalski, J.~Rembielinski and L.~C.~Papaloucas,
  ``Coherent states for a quantum particle on a circle,''
  J.\ Phys.\ A  {\bf 29}, 4149 (1996)
  [quant-ph/9801029].

\bibitem{Gonzalez:1998kj}
  J.~A.~Gonzalez and M.~A.~del Olmo,
  ``Coherent states on the circle,''
  J.\ Phys.\ A  {\bf 31}, 8841 (1998)
  [quant-ph/9809020].



\bibitem{chad}
G. Chadzitaskos, P. Luft and J. Tolar,
``Coherent states on the circle,''
 J.\  Phys.: \ Conf.\  Ser.\ {\bf 284}, 012016 (2011)
[1101.4171[quant-ph]].



\bibitem{Reed:1975uy}
  M.~Reed and B.~Simon,
  {\it Methods of Modern Mathematical Physics. 2. Fourier Analysis, Selfadjointness}, (Accademic Press, NY, 1975).

\bibitem{kbs}
J.~R.~Klauder and B.-S.~Skagerstam,
{\it Coherent States}, (World Scientific, Singapore, 1985).


\bibitem{Klauder:2001ra}
  J.~R.~Klauder,
  ``The current state of coherent states,'' in the Proceddings of the 7th ICSSUR Conference, Boston, MA, June 2001;
  [quant-ph/0110108].

\bibitem{Klauder:2001rb}
  J.~R.~Klauder,
  ``Phase space geometry in classical and quantum mechanics,'' in Contemporary Problems in Mathematical Physics, (World Scientific, Singapore, 2002), pp. 395-408;
  [quant-ph/0112010].

\bibitem{howfar1}
J.~R.~Klauder,
``How far apart are classical and quantum systems?,'' in Squeezed States and Uncertainty Relations (Rinton Press, Princeton, 2003), pp. 180-187;
 [quant-ph/0308049].

\bibitem{howfar}
D.~Abernethy and J.~R.~Klauder,
``The Distance Between Classical and Quantum Systems,''
Foundations of Physics {\bf 35},  881 (2005)
[quant-ph/0411070].

\bibitem{abram}
{\it Handbook of Mathematical Functions},
10th edition,  Appl. Math. Ser. {\bf 55},
Section 15, editors  A. Abramowitz  and  I. A. Stegun,
(Dover, NY, 1972).


\end{thebibliography}
\end{document}